# Structures and Optical Absorption of $Bi_2OS_2$ and $LaOBiS_2$

Akira Miura[1]*, Yoshikazu Mizuguchi[2], Takahiro Takei[3], Nobuhiro Kumada[3], Eisuke Magome[4], Chikako Moriyoshi[4], Yoshihiro Kuroiwa[4], Kiyoharu Tadanaga[1]

[1] *Faculty of Engineering Hokkaido University, Kita 13, Nishi 8, Sapporo 060-8628 Japan*

[2] *Department of Electrical and Electronic Engineering, Tokyo Metropolitan University, 1-1 Minami-osawa, Hachioji, Tokyo 192-0397 Japan*

[3] *Center for Crystal Science and Technology, University of Yamanashi, 7-32 Miyamae, Kofu 400-8511, Japan*

[4] *Department of Physical Science, Hiroshima University, 1-3-1 Kagamiyama, Higashihiroshima, Hiroshima 739-8526 Japan.*

E-mail: amiura@eng.hokudai.ac.jp

The band gaps of isostructural $Bi_2OS_2$ and $LaOBiS_2$ were examined using optical absorption and discussed with the band structures calculated based on the crystal structures determined using synchrotron X-ray diffraction. The Bi 6p and S 3p orbitals in the Bi–S plane were computationally predicted to constitute the bands near the Fermi level. The optical reflectance spectra of $Bi_2OS_2$ and $LaOBiS_2$ showed optical band gaps of ~1.0 eV, which were close to the computationally calculated direct band gaps of ~0.8 eV. Our results show that $Bi_2OS_2$ and $LaOBiS_2$ are semiconductors containing direct band gaps of 0.8–1.0 eV, and they are suggested to be candidates for optoelectronic materials in the near-infrared region without highly toxic elements.

## Introduction

Optoelectronic materials in the near-infrared regions, such as GaAs and CdTe, are important materials for photovoltaics and infrared optoelectronic devices. However, alternative materials with less toxic elements are desired for reducing environmental risks. In this respect, bismuth sulfides/oxysulfides are attractive materials for these applications. For instance, $Bi_2S_3$ is a semiconductor with a direct band gap of 1.3–1.4 eV that is used for solar cells with semiconductor–liquid junctions.[1] $Bi_2O_2S$ is an oxysulfide semiconductor with an indirect gap of approximately 1.12 eV or 1.5 eV, and shows potential for application as photocatalysts and in electro-optics.[2, 3] $Bi_9O_{7.5}S_6$ is a semiconductor with a direct band gap of 1.27 eV. [4]

$MOBiS_2$ (M: Bi, La), which can also be described as $Bi_2OS_2$ and $LaOBiS_2$, are isostructural layered oxysulfides and have been extensively studied because their superconductivity were triggered by substituting $F^-$ for $O^{2-}$.[5, 6] Additionally, recent study has showed the potential of $LaOBiS_2$ as a thermoelectronic material.[7] They consist of alternative stacks of distorted $BiS_2$ rock-salt layers and Bi–O/La–O layers[8, 9] (Fig. 1). Tanryverdiev et al. reported that $LaOBiS_2$ is isostructural with the previously reported $CeOBiS_2$.[8] Recently, the crystal structures of $Bi_2OS_2$ and $LaOBiS_2$ have been analyzed through Rietveld refinement using laboratory X-ray diffraction (XRD) at room temperature with a moderate precision.[9-11] The structural analysis of $LaOBiS_2$ through neutron diffraction at 15 K has also been reported.[12] However, their band gaps are under debate. Band structures predicted using density-functional-theory calculations have showed that $Bi_2OS_2$ and $LaOBiS_2$ have band gaps in the range of 0.4–1.0 eV.[13-17] The semiconductivity of $LaOBiS_2$ single crystals has been reported with its forbidden band of 0.86 eV deduced from a Hall measurement above ~300°C.[8] Nonetheless, a recent report has shown its band gap to be 0.08 eV, which can be attributed to the n-type doping feature in $LaOBiS_2$.[18] Moreover, metallic conductivity has been reported in $Bi_2OS_2$ powder;[10] however, it can be understood by the conductivity of the Bi metal, which was detected through XRD. Therefore, calculating their band structures based on highly accurate crystal structures and optical absorption would help in obtaining a reasonable estimation of their band structures near the Fermi level.

In this report, we investigated the precise crystal structures of $Bi_2OS_2$ and $LaOBiS_2$ through synchrotron powder XRD diffraction and their band structures were computationally calculated. Their optical reflectance spectra were examined for the estimation of their band gaps. The results show that $Bi_2OS_2$ and $LaOBiS_2$ are semiconductors having direct band gaps of 0.8–1.0 eV.

## Experimental and computational

The synthesis of $Bi_2OS_2$ and $LaOBiS_2$ powders were performed from stoichiometric mixtures of oxides and sulfides through solid–solid reactions in glass tubes at 300 and 800 °C, respectively.[5, 6] Synchrotron XRD measurements were performed at room temperature at the BL02B2 beam line at SPring-8 with the approval of 2014A1008, 2014B1071, and 2015A1441. The wavelength of the beam was 0.49575 and 0.49542 Å for the measurements of $Bi_2OS_2$ and $LaOBiS_2$ powders, respectively. Rietveld refinements were performed using RIETAN-FP,[19] and a crystal scheme was drawn using VESTA.[20] Band structures were calculated using the VASP package [21] with PBE-GGA[22] using the crystal structures proposed using Rietveld refinement. The energy cut off was 400 eV. Diffuse reflectance spectra were measured using a V-670 UV-vis-NIR spectrophotometer. A discontinuous gap in absorption at 1.55 eV (800 nm) due to the switch of spectrometers was manually connected by adding constant values to the absorption intensity in the energy region >1.55 eV.

## Results

Figure 2 shows the Rietveld analysis profiles of $Bi_2OS_2$ and $LaOBiS_2$. Main peaks were indexed as tetragonal cells with the space group *P*4/*nmm*. Minor impurity peaks were detected; the $Bi_2OS_2$ profile exhibited 5.5 mass% of $Bi_2S_3$ and 5.4 mass% of Bi, while $LaOBiS_2$ exhibited 6.8 mass% of $La_2O_2S$. The refinement factors, $R_{wp}$, for $Bi_2OS_2$ and $LaOBiS_2$ converged with 3.87% and 5.44%, respectively. The lattice parameters of $Bi_2OS_2$ ($a$ = 3.96086(4) Å, $c$ = 13.8024(2) Å) were shorter than those of $LaOBiS_2$ ($a$ = 4.06456(3) Å, $c$ = 13.86141(15) Å). The selected bond lengths and angles are listed in Table 1. $Bi_2OS_2$ contains a shorter and more zigzag S–Bi–S bond (2.8199(8) Å, 166.63(14)°) than $LaOBiS_2$ does (2.8805(4) Å, 172.36(15)°). Similar bond lengths but slightly flatter Bi–S planes can be found in the isostructural La(O,F)$BiS_2$ single crystals (2.8730(14) Å, 180.8(3)º).[23] Atomic displacement factors of S(1) are larger than those of S(2) in the both $Bi_2OS_2$ and $LaOBiS_2$, which agree with those determined by single-crystal XRD analyses of related layered materials[23-25]. These factors can be understand by large interplane space between Bi–S plains. Large interplane distances between Bi–S plains, which correspond to Bi(1)–S(1) interplane distances, 3.298(8) or 3.448(6) Å, suggest the existence of a lone pair of $Bi^{3+}$.[17] The relatively smaller atomic displacement factors of Bi(1) and S(1) in $Bi_2OS_2$ than those in $LaOBiS_2$ would be attributed to relatively shorter Bi(1)–S(1) interplane distances. The bond length of Bi–O (2.3490(8) Å) in $Bi_2OS_2$ is shorter than the La–O bond in $LaOBiS_2$ (2.3839(6) Å), which may indicate that the interaction between Bi and O is stronger than that between La and O.

Figure 3 shows the calculated band structures of $Bi_2OS_2$ and $LaOBiS_2$ near the Fermi level. An overall similarity was found with the previously calculated band structures.[14-17] The band structures of both $Bi_2OS_2$ and $LaOBiS_2$ are similar near the X and R points, but a difference is found near the Γ and Z points. Near the X and R points, both compounds show similar dispersions of the bands, and direct gaps of 0.82 and 0.84 eV are observed, respectively. This valence-band top and conduction-band bottom are mainly attributed to the S 3p and Bi 6p in the Bi–S(1) plane, respectively.[13-16] Thus, the similarity in the dispersions of the band structures can be attributed to the similar Bi–S planes. Near the Γ and Z points, while $LaOBiS_2$ exhibits relatively localized bands located only a few tens of meV below the Fermi level, $Bi_2OS_2$ displays these at approximately −0.8 eV. As the majority of these bands are composed of O 2p,[15] the lower bands in $Bi_2OS_2$ can be explained by the greater overlap of the O 2p orbital with the Bi 6p orbital.

Band gaps of 0.99 and 1.00 eV are, respectively, derived for $Bi_2OS_2$ and $LaOBiS_2$ using a Tauc plot assuming their direct transition (Fig. 4). $Bi_2S_3$, an impurity of $Bi_2OS_2$, did not significantly affect its optical absorption; the absorption edge, which was observed at 1.29 eV through the reflectance measurement of $Bi_2S_3$ powder (not shown), was not seen. The optical band gap of $Bi_2OS_2$ is also close to the computational prediction (0.82 eV). The optical band gap of $LaOBiS_2$ is similar to those deduced through a Hall measurement (0.86 eV)[8] and the above the computational prediction (0.84 eV). Therefore, a fair degree of accuracy in the estimation of these band structures and optical band gaps is ensured. The similarity in the optical absorbance between $Bi_2OS_2$ and $LaOBiS_2$ can be explained by the similarity in the calculated valence-band top and conduction-band bottom. These band gaps are smaller than those observed in $Bi_9O_{7.5}S_6$ (1.27 eV[4])and $Bi_2O_2S$ (ca. 1.12 eV or 1.5

eV[2, 3]), which may be related to the overlap between the Bi 6p and S 3p as well as the contribution of the Bi 6s and S 3s orbitals.

**Conclusion**

LaOBiS$_2$ and Bi$_2$OS$_2$ have similar Bi–S planes, optical band gaps, and band structures near the Fermi level composed of the Bi 6p and S 3p orbitals, and they show potential for application as optoelectronic materials in the near-infrared region of 0.8–1.0 eV without containing highly toxic elements. The synthesis of thin films of these oxysulfides is desirable for further investigation of their optical absorbance coefficient and optoelectronic properties, and a synthesis temperature of 300$^o$C for Bi$_2$OS$_2$ powder appears attractive for the low-temperature synthesis of its crystalline film.

**Acknowledgments**

A.M. thanks Profs. T. Nakanishi and Y. Hasegawa for optical measurements, and T. Shimada, T. Nagahama, and M. Nagao for discussion. The experiments were partially supported by KAKENHI Grant Numbers 15K14113 and 25707031

## Figure Captions

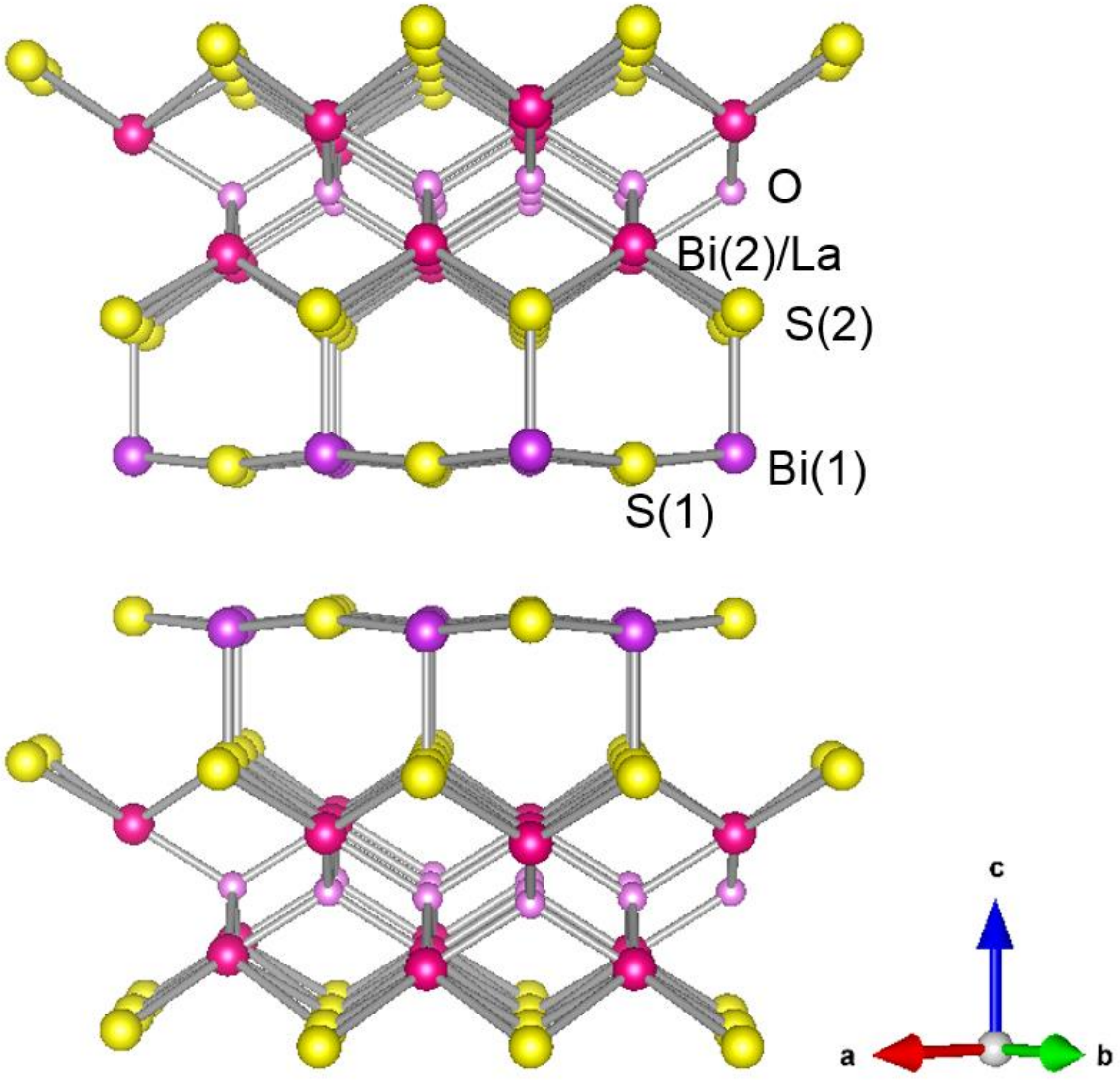

**Fig. 1.** Crystal structures of $Bi_2OS_2$ and $LaOBiS_2$.

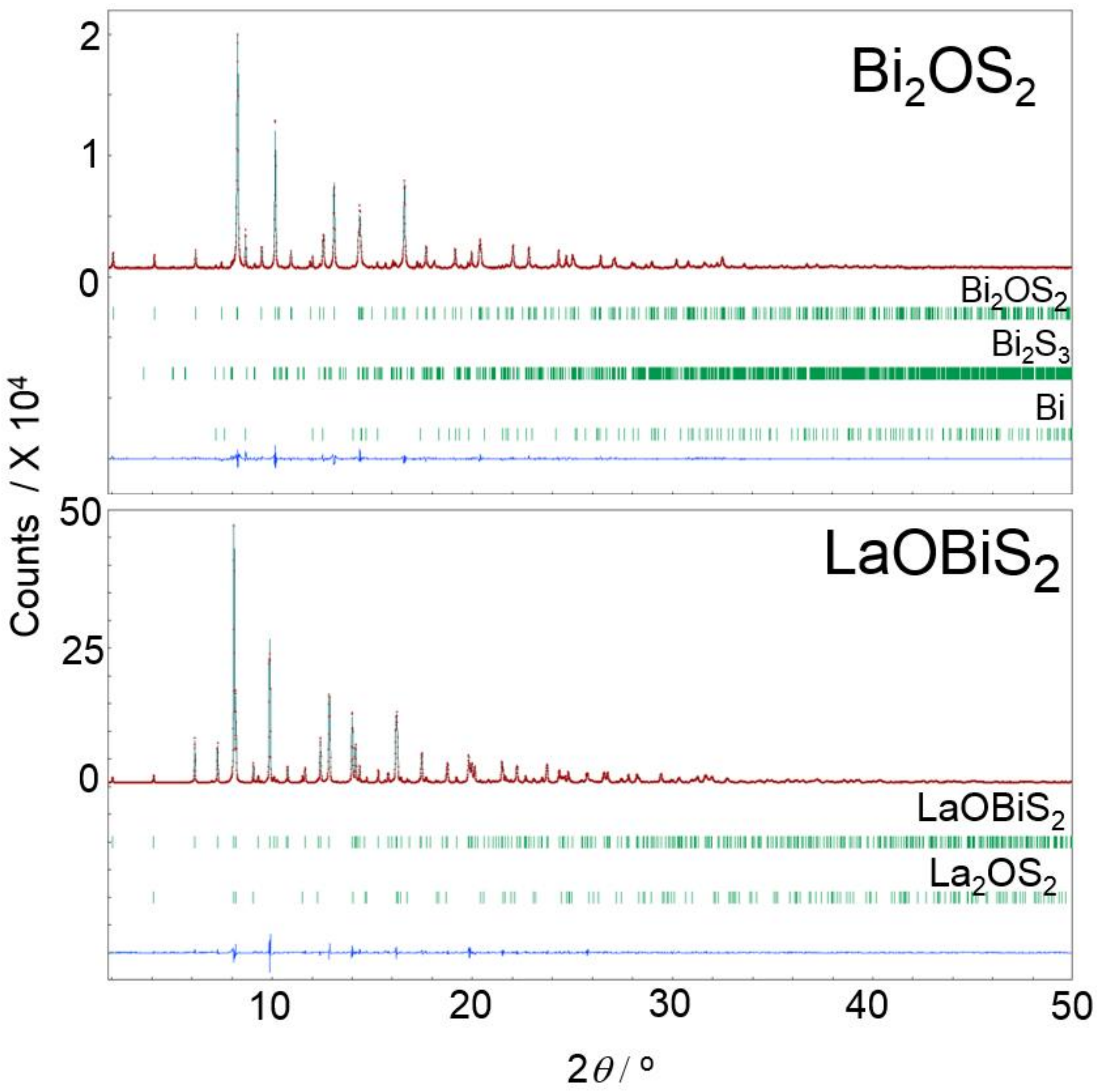

**Fig. 2.** Rietveld profiles of synchrotron X-ray diffraction of $Bi_2OS_2$ and $LaOBiS_2$. Red dots describes experimental points. Residuals are shown in blue line. Green ticks show allowed diffractions.

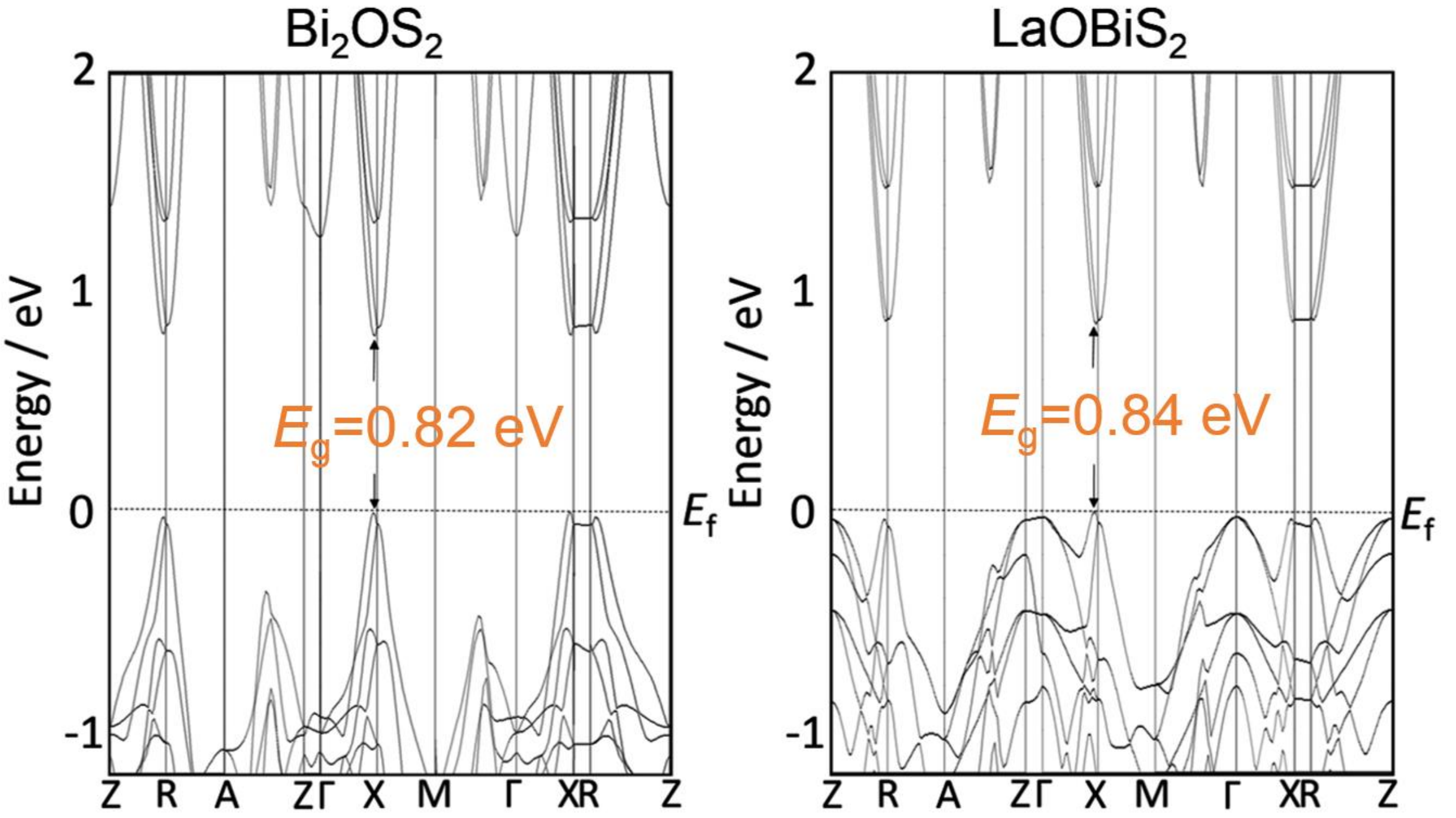

**Fig. 3.** Band structures of $Bi_2OS_2$ and $LaOBiS_2$ calculated using DFT calculation. Fermi level is set to the top of valence band.

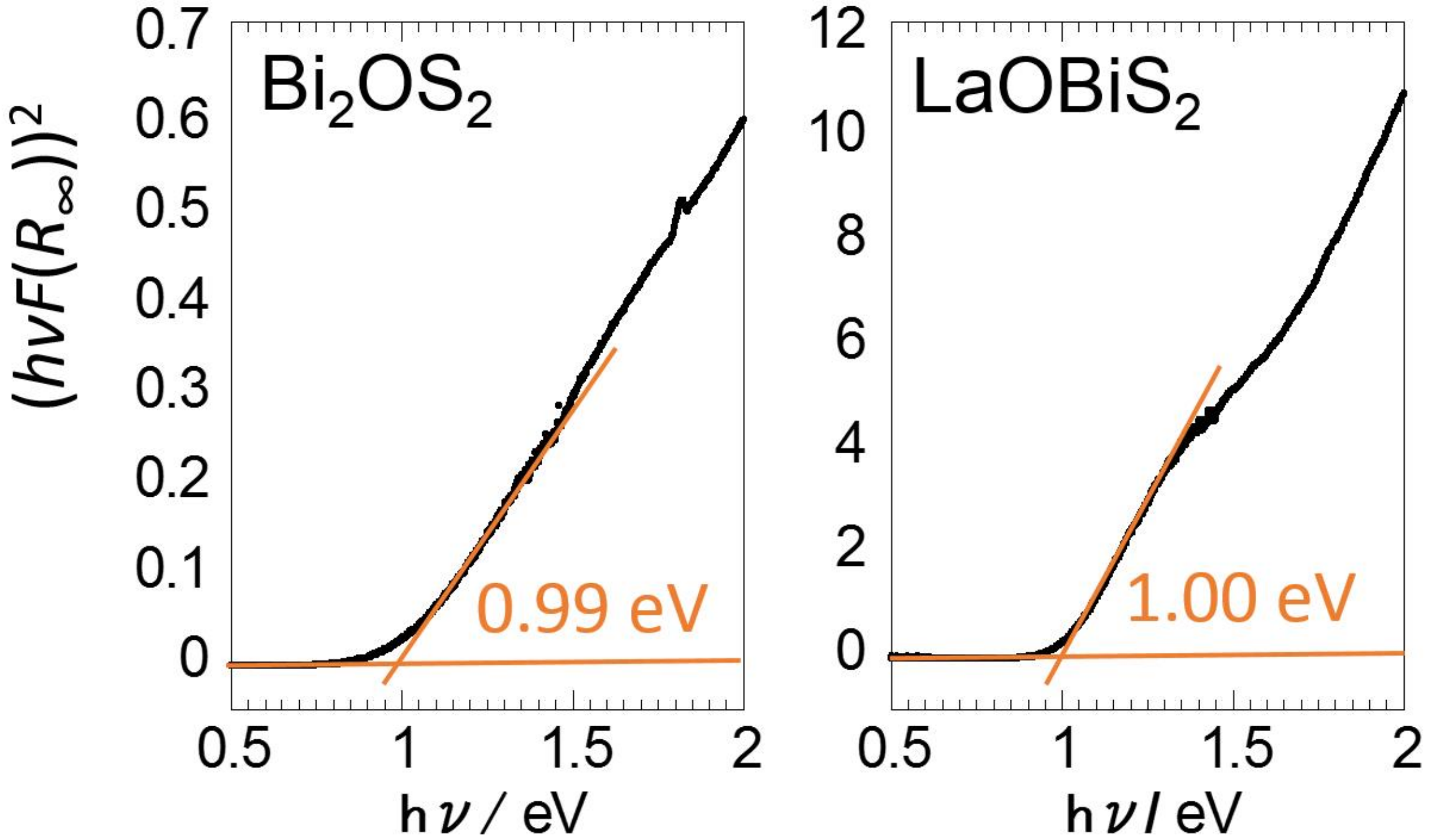

**Fig. 4.** Tauc plots for optical absorption of $Bi_2OS_2$ and $LaOBiS_2$ powders.

**Table I.** Atomic parameters of $Bi_2OS_2$ and $LaOBiS_2$

| Atom | x | y | z | *B* (Å$^2$) |
|---|---|---|---|---|
| | | $Bi_2OS_2$ | | |
| Bi1 | 1/4 | 1/4 | 0.13155(9) | 1.15(3) |
| Bi2 | -1/4 | -1/4 | 0.40823(7) | 1.02(3) |
| S1 | -1/4 | -1/4 | 0.1078(5) | 1.6(2) |
| S2 | 1/4 | 1/4 | 0.3124(5) | 0.92(14) |
| O | -1/4 | -3/4 | 1/2 | 1 |
| | | $LaOBiS_2$ | | |
| Bi1 | 1/4 | 1/4 | 0.13130(8) | 1.01(2) |
| La1 | -1/4 | -1/4 | 0.41010(8) | 0.43(3) |
| S1 | -1/4 | -1/4 | 0.1175(4) | 2.27(15) |
| S2 | 1/4 | 1/4 | 0.3040(3) | 0.51(9) |
| O | -1/4 | -3/4 | 1/2 | 1 |

$Bi_2OS_2$ *a* = 3.96086(4) Å, *c* = 13.8024(2) Å: $LaOBiS_2$ *a*=4.06456(3) Å, *c*=13.86141(15) Å

**Table II.** Selective atomic coordination of $LaOBiS_2$ and $Bi_2OS_2$

| Interatomic | $Bi_2OS_2$ | $LaOBiS_2$ |
|---|---|---|
| Distance (Å) | | |
| Bi(1)–S(1) [in-plane] | 2.8199(8) | 2.8805(4) |
| Bi(1)–S(1) [interplane] | 3.303(7) | 3.448(6) |
| Bi(1)–S(2) | 2.496(7) | 2.476(5) |
| La–O(1) | - | 2.3839(6) |
| Bi(2)–O(1) | 2.3508(6) | - |
| La–S(2) | - | 3.192(3) |
| Bi(2)–S(2) | 3.098(3) | - |
| | | |
| Angle (º) | | |
| S(1)–Bi(1)–S(2) | 96.68(14) | 93.82(13) |
| S(1)–Bi(2)–S(1) | 166.63(16) | 172.36(15) |
| S(2)-La-S(2) | - | 128.43(12) |
| S(2)-Bi(2)-S(2) | 129.43(14) | - |
| La-O-La | - | 116.97(6) |
| Bi(2)-O-Bi(2) | 114.80(6) | - |